\documentclass[prl,twocolumn,showpacs,superscriptaddress,preprintnumbers,amssymb]{revtex4}
\usepackage{graphicx}
\usepackage{dcolumn}
\usepackage{bm}

\newcommand{\beq}{\begin{equation}}
\newcommand{\eeq}{\end{equation}}
\newcommand{\beqn}{\begin{eqnarray}}
\newcommand{\eeqn}{\end{eqnarray}}

\begin{document}

\title{Two dimensional Symmetry Protected Topological Phases with
$\mathrm{PSU}(N)$ \\ and time reversal symmetry}

\author{Jeremy Oon}

\affiliation{Department of Physics, University of California,
Santa Barbara, CA 93106, USA}

\author{Gil Young Cho}

\affiliation{Department of Physics, University of California,
Berkeley, CA 94720, USA}

\author{Cenke Xu}

\affiliation{Department of Physics, University of California,
Santa Barbara, CA 93106, USA}

\begin{abstract}

Symmetry protected topological phase is one type of nontrivial
quantum disordered many-body state of matter. In this work we
study one class of symmetry protected topological phases in two
dimensional space, with both PSU($N$) and time reversal symmetry.
These states can be described by a principal chiral model with a
topological $\Theta-$term. As long as the time-reversal symmetry
and PSU($N$) symmetry are both preserved, the 1+1 dimensional
boundary of this system must be either gapless or degenerate. We
will also construct a wave function of a spin-1 system on the
honeycomb lattice, which is a candidate for the symmetry protected
topological phase with both SO(3) and time-reversal symmetry.

\end{abstract}

\date{\today}

\maketitle

\section{1. Introduction}

The interplay between strong interaction and strong quantum
fluctuation can lead to many remarkable properties in quantum
disordered phases. In the most trivial case, a quantum disordered
phase is fully gapped, and nondegenerate, and its ground state
wave function can be adiabatically connected to a direct product
wave function without any phase transition. These quantum
disordered phases are ``trivial", in the sense that they are
quantum analogues of classical disordered phases, namely they are
completely featureless. The best example of trivial quantum
disordered phase is the Mott insulator phase of spinless bosonic
atoms trapped in an optical lattice. Every state of this Mott
insulator can be adiabatically connected to direct product state $
\prod_i | \hat{n}_i = k \rangle_i $, where $\hat{n}_i$ is the
boson number operator on site $i$, and $k$ is an integer.

The description of trivial quantum disordered phases is
semiclassical, $i.e.$ we can describe the trivial quantum
disordered phase using a field theory defined with a Landau order
parameter. For example, a trivial quantum disordered phase of a
system with SO(3) spin rotation symmetry can be described by a
semiclassical nonlinear sigma model (NLSM) defined with the
N\'{e}el order vector $\vec{n}$ with unit length $(\vec{n})^2 =
1$: \beqn \mathcal{S} = \int d^dx d\tau \ \frac{1}{g}
(\partial_\mu \vec{n})^2. \label{o3}\eeqn In spatial dimensions
higher than 1, by tuning the parameter $g$, there is an
order-disordered phase transition: when $g < g_c$, $\vec{n}$ is
ordered, and the SO(3) symmetry is spontaneously broken; when $g >
g_c$, $\vec{n}$ is disordered, and the ground state wave function
of this quantum disordered phase is approximately a direct product
state $| 0 \rangle \sim \prod_i | l = 0\rangle_i $, where $l$ is
the angular momentum quantum number on every site.

Now there is a consensus that, quantum disordered phases of
strongly correlated systems can have much richer structures
compared with classical disordered phases. Roughly speaking, three
types of ``nontrivial" quantum disordered phases have been
studied: ({\it i.}) topological phases, whose ground state is
fully gapped but
topologically degenerate; 
({\it ii.}) Algebraic spin (Bose) liquid phase, which is still a
quantum disordered phase of a bosonic spin system, but the
spectrum remains gapless, and the physical quantities have a
power-law instead of short range correlation; ({\it iii.})
Symmetry protected topological (SPT) phases, whose bulk spectrum
is identical to a trivial quantum disordered phase, but its
boundary must be either gapless or degenerate when and only when
the system has certain symmetry $G$. The ground state wave
function of a SPT is completely different and cannot be
continuously connected to a trivial wave function without a phase
transition, when the Hamiltonian is invariant under symmetry $G$.
The 2d quantum spin Hall insulator and the 3d topological
insulator are both SPT phases with time-reversal symmetry.

In this work we will focus on SPT phases of bosonic spin systems.
SPT is a pure quantum phenomenon, and there is no analogue in
classical world. Thus one would expect that the description for
these states should be purely quantum, and a semiclassical
formalism should completely fail to describe them. However, we
will try to demonstrate that, the SPT phases can still be
described semiclassically using NLSMs like Eq.~\ref{o3}, as long
as we include an appropriate topological term. We want to stress
that, in our approach, the target space of the NLSM is the
manifold of a semiclassical order parameter. This is different
from the approach in Ref.~\cite{wenspt,wenliu}, where a NLSM was
also introduced to describe SPT phases, but the target space of
this NLSM is the group manifold of the symmetry $G$, instead of
the manifold of an order parameter.

At least in one dimensional systems, semiclassical NLSMs have been
proved successful in describing SPT phases. For example, the O(3)
NLSM in Eq.~\ref{o3}, plus a topological $\Theta-$term describes a
spin-1 Heisenberg chain when $\Theta = 2\pi$, and it is well-known
that the spin-1 antiferromagnetic Heisenberg model is a SPT phase
with 2-fold degeneracy at each boundary~\cite{haldane1,haldane2}.
This two fold degeneracy at the boundary can be read off from this
1d O(3) NLSM, since its boundary is a 0+1d O(3) NLSM with a
Wess-Zumino-Witten term at level $ k = 1 $, whose ground state is
two fold degenerate~\cite{ng1994}.

In principle, a NLSM describes a system with long correlation
length. Thus a NLSM plus a $\Theta-$term most precisely describes
a SPT phase tuned {\it close to} a critical point (but still in
the SPT phase). When a SPT phase is tuned close to a critical
point, the NLSM not only describes its topological properties
($e.g.$ edge states $etc.$), but also describes its dynamics, for
example excitation spectrum above the energy gap (much smaller
than the ultraviolet cut-off). When the SPT phase is tuned deep
inside the SPT phase, namely the correlation length is comparable
with the lattice constant, this NLSM can no longer describe its
dynamics accurately, but since the topological properties of this
SPT phase is unchanged while tuning, these topological properties
can still be described by the NLSM.

In Ref.~\cite{xu3dspt}, the author discussed a class of three
dimensional SPT phases with SU($N$) symmetry (Rigorously speaking,
the symmetry of the SPT constructed in Ref.~\cite{xu3dspt} is
PSU($N$) = SU($N$)$/Z_N$, where $Z_N$ is the center of SU($N$)),
and just like the Haldane phase in 1d, these 3d SPT phases are
described by a NLSM defined with the N\'{e}el order parameter
only. For SU($N$) magnet, the manifold $\mathcal{M}$ of the
N\'{e}el order is $\mathcal{M} =
\frac{\mathrm{U}(N)}{\mathrm{U}(m)\times \mathrm{U}(N -
m)}$~\cite{sachdev1989nucl,sachdev1990}. When $N
> 2$ and $0 < m < N$, $\pi_4[\mathcal{M}] = \mathbb{Z}$. Thus a
nontrivial topological $\Theta-$term can be defined for the
SU($N$) N\'{e}el order parameter, and when $\Theta = 2\pi$, it was
argued that the system is a three dimensional SPT, whose 2+1d
boundary must be either gapless or degenerate~\cite{xu3dspt}.

In 2 dimensional space, a straightforward generalization of
Ref.~\cite{xu3dspt} is difficult, since for $N > 2$,
$\pi_3[\frac{\mathrm{U}(N)}{\mathrm{U}(m)\times \mathrm{U}(N -
m)}] = \mathbb{Z}_1$, thus a $\Theta-$term is not well defined for
manifold $\mathcal{M}$ in two dimensions with general $N$. For the
simplest case with $N = 2$ and $m = 1$, $i.e.$ the ordinary SU(2)
N\'{e}el order whose manifold is $S^2$, although the homotopy
group $\pi_3[S^2] = \mathbb{Z}$, a nontrivial mapping from the
space-time manifold to the target space $S^2$ cannot be written as
an integral of local terms of the N\'{e}el order parameter, thus
it is much more complicated. Alternatively, in this paper we will
study 2d systems with symmetry $\mathrm{PSU}(N) \times
\mathbf{Z}^T_2$,
where $\mathbf{Z}^T_2$ is the time-reversal symmetry. 
We will demonstrate that for a system with $\mathrm{PSU}(N) \times
\mathbf{Z}^T_2$ symmetry, a topological $\Theta-$term can be
defined in 2+1 dimensional space-time with semiclassical order
parameters, and when $\Theta = 2\pi$ the topological term will
drive the system into a SPT phase.

\section{2. Field theory description}

\subsection{2.1 SPT phase at $\Theta = 2\pi$}

In this section we will discuss the field theory description of
the 2+1d SPT phase with $\mathrm{PSU}(N)\times \mathbf{Z}^T_2$
symmetry. As we discussed in the introduction, we will take the
semiclassical formalism, and define the field theory in terms of
the order parameter of PSU(N), whose configurations form manifold
$\mathcal{M} = \frac{\mathrm{U}(N)}{\mathrm{U}(m)\times
\mathrm{U}(N - m)}$. Every element $\mathcal{P}$ of the manifold
$\mathcal{M}$ can be represented as \beqn \mathcal{P} = V^\dagger
\Omega V, \ \ \ \Omega = \left(
\begin{array}{cccc}
\mathbf{1}_{m \times m}, & \mathbf{0}_{m \times N - m} \\ \\
\mathbf{0}_{N - m \times m}, & \mathbf{-1}_{N - m \times N - m}
\end{array}
\right), \eeqn where $V$ is a SU($N$) transformation matrix.
$\mathcal{P}$ is a Hermitian matrix that satisfies $\mathcal{P}^2
= \mathbf{1}_{N \times N}$. In fact, when $N = 2$, $\mathcal{M}$
is precisely $S^2$. The matrix order parameter $\mathcal{P}$ is
always invariant under the center $Z_{N}$ of SU($N$): $V = e^{i 2
\pi k /N} \mathbf{1}_{N \times N}$ ($k = 1, \cdots N -1$), thus a
NLSM defined with $\mathcal{P}$ has symmetry $\mathrm{PSU}(N) =
\mathrm{SU}(N)/Z_N$. When $N = 2$, PSU(2)$=$SO(3) is the ordinary
spin rotation group of model Eq.~\ref{o3}. If the symmetry of a
quantum spin system is SO(3) instead of SU(2), then the Hilbert
space on every site of the system must be a representation of
SO(3), thus we are restricted to integer spin systems only. With
integer spins, $(\mathbf{Z}^T_2)^2 = +1$.

For a general $N$, The homotopy group $\pi_3[\mathcal{M}] =
\mathbb{Z}_1$, thus a NLSM defined with $\mathcal{P}$ does not
have a nontrivial topological $\Theta-$term in 2+1d space-time.
Thus for a 2+1d system with PSU($N$) symmetry only, there is no
nontrivial SPT that can be described using semiclassical order
parameter $\mathcal{P}$. Now let us combine PSU($N$) and
time-reversal symmetry together, and define the following order
parameter $U$: \beqn U = \cos(\theta) \mathbf{1}_{N \times N} + i
\sin(\theta) \mathcal{P}. \label{order}\eeqn Now $U$ is a unitary
matrix, and $U \in \mathrm{U}(N)$. Since $\pi_3[\mathrm{U}(N)] =
\mathbb{Z}$, a principal chiral nonlinear sigma model can be
defined with order parameter $U$, plus a nontrivial topological
$\Theta-$term: \beqn \mathcal{S} &=& \int d^2x d\tau \ \frac{1}{g}
\mathrm{tr}[\partial_\mu U^\dagger
\partial_\mu U] \cr\cr &+& \frac{i\Theta}{24\pi^2}
\mathrm{tr}[U^\dagger \partial_\mu U U^\dagger \partial_\nu U
U^\dagger \partial_\rho U] \epsilon_{\mu \nu \rho}.
\label{pcm}\eeqn In the simplest case with $N = 2$ and $m = 1$,
the order parameter $\mathcal{P}$ is equivalent to an O(3) vector
$\vec{n}$: $\mathcal{P} = \vec{n}\cdot \vec{\sigma}$. Then
Eq.~\ref{pcm} is equivalent to an O(4) NLSM: \beqn \mathcal{S} &=&
\int d^2x d\tau \ \frac{1}{g} (\partial_\mu \vec{\phi})^2 \cr\cr
&+& \frac{i\Theta}{12\pi^2} \epsilon_{abcd} \epsilon_{\mu \nu
\rho} \phi^a \partial_\mu \phi^b \partial_\nu \phi^c \partial_\rho
\phi^d, \label{o4theta} \eeqn where $\vec{\phi} = (\cos(\theta), \
\sin(\theta) \vec{n})$.

The matrix $U$ has a SU($N$)$_{\mathrm{left}}$ transformation and
a SU($N$)$_{\mathrm{right}}$ transformation, but there are higher
order terms in the Eq.~\ref{pcm} that break the two SU($N$)
symmetries down to its diagonal subgroup PSU($N$).
For general $N$, we can define the following transformations:
\beqn \mathrm{SU}(N) &:& \mathcal{P} \rightarrow V^\dagger
\mathcal{P} V, \cr \cr \mathbf{Z}_2^T &:& \mathcal{P} \rightarrow
\mathcal{P}^\ast, \ \ \ \theta \rightarrow \pi - \theta, \ \ \ V
\rightarrow V^\ast. \eeqn Under this definition, the SU($N$) and
$\mathbf{Z}^T_2$ transformations commute with each other.

In the follows, we will argue that, when $\Theta = 2\pi$,
Eq.~\ref{pcm} with symmetry SU($N$)$\times \mathbf{Z}^T_2$
describes a SPT, whose 1+1d boundary must be either gapless or
degenerate.

In Eq.~\ref{pcm}, by tuning $g$, there is an order-disorder phase
transition. We will always focus on the disordered phase of
Eq.~\ref{pcm}, thus we will focus on the phase with a large
coupling constant $g$. When $\Theta = 2\pi$, the bulk spectrum of
the quantum disordered phase is identical to the case with $\Theta
= 0$, thus the bulk is fully gapped and nondegenerate. Then one
can safely integrate out the bulk and focus on the boundary. At
$\Theta = 2\pi$, the boundary of the system is described by the
following principal chiral model with a Wess-Zumino-Witten (WZW)
term: \beqn \mathcal{S}_b &=& \int dx d\tau \ \frac{1}{g}
\mathrm{tr}[\partial_\mu U^\dagger
\partial_\mu U] \cr\cr &+& \int dx d\tau du \ \frac{i 2\pi}{24\pi^2}
\mathrm{tr}[U^\dagger \partial_\mu U U^\dagger \partial_\nu U
U^\dagger \partial_\rho U] \epsilon_{\mu \nu \rho}. \label{bpcm}
\eeqn Here $U(x, \tau, u)$ is an extension of physical order
parameter $U(x, \tau)$ that satisfies \beqn U(x, \tau, u = 0) &=&
\mathbf{1}_{N \times N}, \cr\cr U(x, \tau, u = 1) &=& U (x, \tau).
\eeqn For the simple case with $N = 2m = 2$, Eq.~\ref{bpcm}
reduces to a 1+1d O(4) NLSM with a WZW term at level-1: \beqn
\mathcal{S} & = & \int dx d\tau \ \frac{1}{g} (\partial_\mu
\vec{\phi})^2 \cr\cr &+& \int dx d\tau du \ \frac{2\pi i}{12\pi^2}
\epsilon_{abcd} \epsilon_{\mu \nu \rho} \phi^a \partial_\mu \phi^b
\partial_\nu \phi^c
\partial_\rho \phi^d, \label{o4wzw} \eeqn

This principal chiral model Eq.~\ref{bpcm}, with a full
SU($N$)$_\mathrm{left} \times$SU($N$)$_\mathrm{right}$ symmetry,
was proved to be a gapless conformal field
theory~\cite{witten1984,KnizhnikZamolodchikov1984}. However, in
our system the SU($N$)$_\mathrm{left}
\times$SU($N$)$_\mathrm{right}$ symmetry is broken down to the
diagonal SU($N$), thus the conformal field theory might be gapped
out due to relevant perturbations introduced by this symmetry
reduction. With this symmetry reduction, the boundary theory
Eq.~\ref{bpcm} is reduced to the following NLSM with a
$\Theta^\prime-$term: \beqn \mathcal{S}_b = \int dx d\tau \
\frac{1}{g} \mathrm{tr}[(\partial_\mu \mathcal{P})^2] + \frac{
\Theta^\prime }{16 \pi} \mathrm{tr}[ \mathcal{P} \partial_\mu
\mathcal{P}
\partial_\nu \mathcal{P} ] \epsilon_{\mu \nu}, \label{thetap}
\eeqn as long as the $\mathbf{Z}^T_2$ symmetry is preserved,
namely the expectation value of $\cos(\theta)$ is zero, the
boundary $\Theta^\prime$ is fixed at $\Theta^\prime = \pi$. Under
the $\mathbf{Z}^T_2$ transformation, $\Theta^\prime \rightarrow
2\pi - \Theta^\prime$. If the time-reversal symmetry is explicitly
broken, namely a background field that couples linearly to
$\cos(\theta)$ is turned on, then at the boundary $\Theta^\prime$
is also tuned away from $\pi$: $\Theta^\prime = 2\theta -
2\cos(\theta)\sin(\theta)$.

If we ignore the physical interpretation of the field
$\mathcal{P}$, this 1+1d NLSM at $\Theta^\prime = \pi$
(Eq.~\ref{thetap}) can be used to describe the SU($N$)
antiferromagnet with conjugate representations on A and B
sublattices~\cite{affleck1985,sachdev1989nucl,sachdev1990}, and
$\mathcal{P}$ is precisely the SU($N$) N\'{e}el order parameter.
In fact, for the simplest case with $N = 2$, $m = 1$,
Eq.~\ref{thetap} precisely reduces to an O(3) NLSM with a
$\Theta^\prime$ term with $\Theta^\prime = \pi$: \beqn \mathcal{S}
= \int dx d\tau \ \frac{1}{g} (\partial_\mu \vec{n})^2 +
\frac{i\Theta^\prime }{8\pi} \epsilon_{abc}\epsilon_{\mu\nu} n^a
\partial_\mu n^b \partial_\nu n^c. \label{o3theta}\eeqn
With $\Theta^\prime = \pi$, this model describes an
antiferromagnetic spin-1/2 chain, and based on the
Lieb-Schultz-Mattis (LSM) theorem this theory must be either
gapless or degenerate~\cite{LSM}.

In Eq.~\ref{thetap}, $\Theta^\prime = \pi$ is the transition point
between two stable fixed points at $\Theta^\prime = 0$ and $2\pi$,
and this transition is driven by tuning $\Theta^\prime$. For
example, when $m = 1$, Eq.~\ref{thetap} becomes the CP$^{N-1}$
model with $\Theta^\prime = \pi$, and in the large$-N$ limit it
was demonstrated that the CP$^{N-1}$ model has a first order
transition at $\Theta^\prime =
\pi$~\cite{affleck1985,coleman1976}. For general $m$ and $N$, this
transition at $\Theta^\prime = \pi$ was analyzed through a
renormalization group calculation of both $g$ and $\Theta^\prime$
as in
Ref.~\cite{pruisken1,pruisken2,pruisken3,pruisken4,pruisken2011}.
If this transition is continuous, then this 1+1d boundary system
must be a gapless CFT at $\Theta^\prime = \pi$; if this transition
is first order, then this boundary system must be two fold
degenerate at $\Theta^\prime = \pi$~\cite{xuludwig}. Thus we
conclude when the bulk theory Eq.~\ref{pcm} has $\Theta = 2\pi$,
its boundary must be nontrivial, $i.e.$ it must be either gapless
or degenerate.

To demonstrate that this phase is a SPT, we need to show that its
boundary can only be realized as the boundary of a 2+1d system,
$i.e.$ it cannot be realized as a real one dimensional lattice
quantum spin system with the same symmetry. For example, the
boundary of a 2d quantum spin Hall insulator is a 1d helical
Luttinger liquid with central charge 1, and it can be argued that
this 1d helical Luttinger liquid cannot be realized as a 1d
electron system with time-reversal symmetry~\cite{wuedge}. Also,
the boundary of 3d topological insulator is a single 2d Dirac
cone, which cannot be realized in a pure 2d system with
time-reversal symmetry. Let us take the simplest case with $N = 2m
= 2$ as an example. In order to argue that Eq.~\ref{o3theta}
cannot be realized in a 1d system with SO(3)$\times
\mathbf{Z}^T_2$ symmetry, let us break the time-reversal symmetry
at the boundary, but make a domain wall of the time-reversal
symmetry breaking pattern: \beqn \Theta^\prime > \pi, \
\mathrm{for} \ x > 0; \ \ \ \ \Theta^\prime < \pi, \ \mathrm{for}
\ x < 0. \label{domainwall} \eeqn The two sides of the domain wall
are conjugate to each other under time-reversal transformation.
Based on the renormalization group calculation of
Ref.~\cite{pruisken1,pruisken2,pruisken3,pruisken4,pruisken2011},
and the argument in Ref.~\cite{xuludwig}, when $\Theta^\prime >
\pi$ the system is in the same phase as $\Theta^\prime = 2\pi$,
while when $\Theta^\prime < \pi$ the system is in the same phase
as $\Theta^\prime = 0$, both phases are fully gapped and
nondegenerate. At the domain wall this system is described by a
0+1d O(3) NLSM with a WZW term at level-1: \beqn \mathcal{S}_0 =
\int d\tau \ \frac{1}{g} (\partial_\mu \vec{n})^2 + \int d\tau du
\ \frac{2\pi i}{8\pi} \epsilon_{\mu\nu} \epsilon_{abc} n^a
\partial_\mu n^b \partial_\nu n^c, \eeqn which is precisely the model
describing a single spin-1/2. A spin-1/2 excitation is not a
representation of SO(3) (it is a representation of SU(2)), since
it is not invariant under the center of SU(2). Also, under the
square of time-reversal transformation, the wave function of a
spin-1/2 excitation would change sign: $(\mathbf{Z}^T_2)^2 =  -1$.
This implies that physical symmetries fractionalize at the domain
wall.


Although one dimensional spin chains can have fractionalized
excitations, this phenomenon of deconfined domain-wall
fractionalization cannot happen in a one dimensional integer spin
chain. Consider for example two 1d systems with integer spins and
SO(3) symmetry, then if these two systems are conjugate to each
other under time-reversal symmetry, then they must be either both
1d SPT, or both trivial states. Then at their domain wall there
should be either an integer localized domain wall spin, or no
domain wall spin at all.

The analysis of domain wall state at the boundary can be
generalized to arbitrary $N$. For arbitrary $N$, the theory
Eq.~\ref{pcm} is parametrized by integer $m$. For general $N$ and
$m$, a single spin with representation Fig.~\ref{monolattice}$b$
is localized at the domain wall Eq.~\ref{domainwall}, and the
theories with different $m$ (mod $N$) have inequivalent domain
wall spins. This domain wall state is always a fractionalized
excitation of PSU($N$), since it is not invariant under the center
of SU($N$). Based on all the analysis above, we can conclude that
Eq.~\ref{pcm} with $\Theta = 2\pi$ is a SPT with symmetry
PSU($N$)$\times \mathbf{Z}^T_2$. Our formalism suggests that for
general $N$, there are {\it at least} $N$ inequivalent phases:
there is one trivial phase, and $N - 1$ SPT phases described by
Eq.~\ref{pcm} with $m = 1, \cdots N -1$, which have different
localized spins at the domain wall Ref.~\ref{domainwall}. For the
simplest case with $N = 2m = 2$, $i.e.$ the case with SO(3)$\times
\mathbf{Z}_2^T$ symmetry, there are only two inequivalent phases,
which is consistent with the classification in Ref.~\cite{wenspt}.
For general $N$, our prediction can be compared with future group
cohomology computation.

\subsection{2.2 Physics with $\Theta = \pi$}

\begin{figure}
\includegraphics[width=3.2 in]{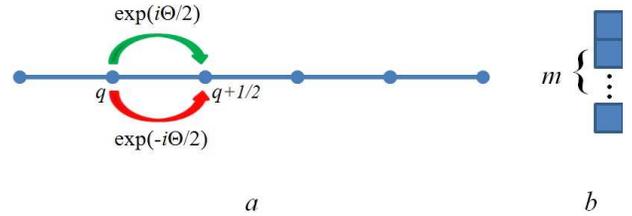}
\caption{($a$) We map Eq.~\ref{Philag} to a one dimensional
tight-binding model. Hopping between nearest neighbor sites
corresponds to changing $\Phi$ by $1/2$. The $\Theta$-term grants
two types of monopoles a factor $\exp(i\Theta/2)$ and $\exp(-
i\Theta/2)$ respectively, which forbids nearest neighbor hopping
when $\Theta = \pi$ due to destructive interference between these
two types of monopoles. ($b$) The SU($N$) spin localized at the
domain wall Eq.~\ref{domainwall}.} \label{monolattice}
\end{figure}

It was first discussed in Ref.~\cite{senthilfisher} that, the O(4)
NLSM Eq.~\ref{o4theta} at $\Theta = \pi$ cannot be trivially
gapped without degeneracy. In Ref.~\cite{xuludwig}, using a
different argument, it was concluded for general principal chiral
models Eq.~\ref{pcm} that their quantum disordered phases must be
either gapless or two fold degenerate when $\Theta = \pi$. In
Ref.~\cite{xuludwig} it was assumed that the system has a full
SU($N$)$_{\mathrm{left}} \times$SU($N$)$_\mathrm{right}$ symmetry.
In our current case, the actual symmetry is PSU($N$)$\times
\mathbf{Z}^T_2$. In this section, using a different argument from
Ref.~\cite{xuludwig}, we will make the same conclusion for
Eq.~\ref{pcm}, $i.e.$ its quantum disordered phase cannot be
gapped without degeneracy when $\Theta = \pi$.

In order to argue that a system is either gapless or degenerate
when $\Theta = \pi$, we only need to argue that if the system is
gapped, it must be degenerate. Thus let us assume it is gapped in
the disordered phase of $\Theta = \pi$. Under this assumption, the
coupling constant $g$ must flow to infinity in the infrared limit
under renormalization group, this is because if $g$ flows to any
fixed point with finite constant $g^\ast$, then the system must be
scaling invariant and gapless. Thus $g$ must flow to infinity once
we assume the system is gapped. Now let us take $g$ to infinity,
then the first term of Eq.~\ref{pcm} vanishes, and Eq.~\ref{pcm}
reduces to a pure topological $\Theta-$term: \beqn \mathcal{S} =
\int d^2x d\tau \ \frac{i\Theta}{24\pi^2} \mathrm{tr}[U^\dagger
\partial_\mu U U^\dagger \partial_\nu U U^\dagger \partial_\rho U]
\epsilon_{\mu \nu \rho}. \label{pcm2}\eeqn

This $\Theta-$term contributes phase factor $\exp(i\Theta)$ to
every SU($N$) instanton in the space-time. However, since our
system only has PSU$(N)\times \mathbf{Z}^T_2$ symmetry instead of
a full SU($N$)$\times$SU($N$) symmetry, an instanton will
fractionalize into two monopoles. A monopole centered at the
origin has the following configuration: \beqn U &=&
\cos(\theta)\mathbf{1}_{N \times N} + i\sin(\theta)\mathcal{P},
\cr\cr && \theta(|\vec{R}| = 0) = 0, \ \mathrm{or} \ \pi, \ \ \
\theta(|\vec{R}| = \infty) = \pi/2; \cr\cr && \int_{|\vec{R}| = R}
d^2R \ \frac{i}{16\pi} \mathrm{tr}[\mathcal{P} \partial_\mu
\mathcal{P}
\partial_\nu \mathcal{P}]\epsilon_{\mu\nu} = 1. \eeqn Here
$\vec{R}$ is the coordinate in the 2+1d Euclidean space-time. In
the simplest case with $N = 2m = 2$, this monopole is the ordinary
``hedgehog" monopole of vector $\vec{n}$ in the space-time. When
$\theta = 0$ and $\pi$ at the origin $|\vec{R}| = 0$, this
monopole carries instanton number $1/2$ and $-1/2$ respectively,
thus this monopole contributes phase factor $\exp(\pm i\Theta/2)$
to the partition function.

Since we are interested in the bulk physics, we can compactify the
two dimensional space into a sphere $S^2$. Now let us define the
following quantity $\Phi(\tau)$ for every time slice $\tau$: \beqn
\Phi(\tau) = \int d^2x \ \frac{i}{32\pi} \mathrm{tr}[\mathcal{P}
\partial_i \mathcal{P}
\partial_j \mathcal{P}]\epsilon_{ij}, \eeqn and since $\pi_2[\mathcal{M}] =
\mathbb{Z}$, $\Phi(\tau)$ is quantized as integer or half-integer,
as long as the configuration of $\mathcal{P}$ has no singularity
at time $\tau$. $\Phi(\tau)$ is increased and decreased through
the monopoles described in the previous paragraph, and one
monopole in the space-time changes $\Phi$ by $1/2$: $\Phi(\tau = +
\infty) - \Phi(-\infty) = n_m/2$, where $n_m$ is the monopole
number in the 2+1d space-time.

Now under the assumption that the system is gapped (hence $g$
flows to infinity), Eq.~\ref{pcm} and Eq.~\ref{pcm2} reduce to the
following single particle quantum mechanics problem defined on a
periodic lattice with lattice constant $1/2$
(Fig.~\ref{monolattice}$a$): \beqn S = \int d\tau \ \frac{1}{2} m
(\partial_\tau \Phi)^2 \pm i \Theta
\partial_\tau \Phi + V(\Phi), \ \ m \rightarrow 0,
\label{Philag} \eeqn where $V(\Phi)$ is a deep periodic potential
that makes $\Phi$ takes only integer and half-integer values. Thus
the original principal chiral model has now reduced to a single
particle tight binding model, where each lattice site corresponds
to a quantized value of $\Phi$. Hopping from site $q$ to site $q +
1/2$ corresponds to a monopole in the space-time, and there are
two different types of monopole, depending on the sign of
$\cos(\theta)$ at the monopole core. The $\Theta$-term will
contribute a factor $\exp(i\Theta/2)$ and $\exp(- i\Theta/2)$ to
the monopole with $\cos(\theta) = +1$ and $\cos(\theta) = -1$ at
the core respectively. With time-reversal symmetry, the two types
of monopoles are degenerate, thus when $\Theta = \pi$ these two
types of monopoles have complete destructive interference with
each other, $i.e.$ hopping by one lattice constant in
Fig.~\ref{monolattice}$a$ is forbidden. However, hopping by two
lattice constants is still allowed, but the band structure will be
{\it doubly degenerate}, namely on this one dimensional lattice
(Fig.~\ref{monolattice}$a$) the states with lattice momentum $p =
0$ and $p = 2\pi$ are degenerate.



If $\Theta$ is tuned away from $\pi$, then the nearest neighbor
hopping in the tight-binding model Eq.~\ref{Philag} is allowed,
and the ground state of Eq.~\ref{Philag} is nondegenerate. Now we
have argued that once the system is gapped at $\Theta = \pi$, it
must be two fold degenerate, namely the system must be either
gapless or degenerate at $\Theta = \pi$. The analysis in this
section implies that when we tune $\Theta = 2\pi$ to 0, the bulk
spectrum must change at $\Theta = \pi$, $i.e.$ the SPT phase and
trivial phase must be separated by a bulk transition at $\Theta =
\pi$.

\section{3. Lattice construction for SPT with
$\mathrm{SO}(3)$$\times \mathbf{Z}^T_2$ symmetry}

\begin{figure}
\includegraphics[width=1\columnwidth]{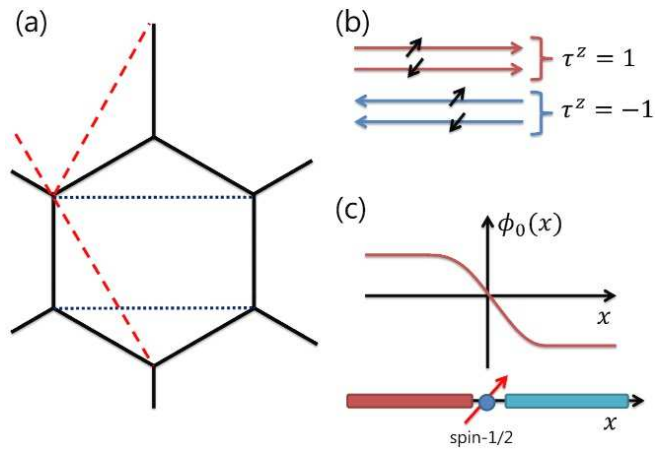}
\caption{Lattice construction for $N=2, m=1$ and its edge state.
(a) the mean field Hamiltonian Eq.~\ref{mf2} consists of the
nearest-neighbor hopping $t$ and the ``color"-orbit interaction
$\lambda$ and $t'$. The dashed lines represent the usual
``color"-orbit coupling $\sim \lambda$ in the Kane-Mele model. It
is equivalent to the Kane-Mele model except for an additional
$x$-directional ``color"-orbit couplings $\sim t'$ (represented by
the dotted lines) on top of the usual ``color"-orbit term $\sim
\lambda$. This anisotropic coupling breaks the gauge symmetry down
to SU(2) and the lattice rotational symmetry which is irrelevant
for the symmetry protected phase. (b) the edge theory consists of
right-moving ($\tau^{z} = 1$) and left-moving ($\tau^{z} = -1$)
spin-$1/2$ doublets. The time-reversal symmetry maps a right-mover
into a left-mover with the {\it opposite} spin state, $i.e.$ the
color and the spin are both flipped by the time-reversal symmetry.
(c) the domain wall in $\phi_{0}(x)$ Eq.(24) traps a localized
spin-$1/2$. This can be derived from the WZW term written in terms
of $(\phi_{0}, \phi_{\mu}), \mu = 1,2,3$ as discussed below
Eq.~\ref{o3theta}. } \label{Fig}
\end{figure}

In this section we will try to construct a lattice spin state for
the 2d SPT with SO(3) and time-reversal symmetry. Since the
Hilbert space on every site must be a representation of SO(3)
group, we must start with an integer spin system. Let us consider
a spin-1 system on a honeycomb lattice, and we will construct a
spin many-body wave function using the following two-color slave
fermion formalism, which was introduced to understand the spin
liquid phenomena observed in Ba$_3$NiSb$_2$O$_9$~\cite{xuspin1}:
\begin{eqnarray} \hat{S}^\mu_i = \frac{1}{2} \sum_{\alpha, \beta =
\uparrow, \downarrow} \sum_{a = 1, 2} f^\dagger_{\alpha, a, i}
\sigma^\mu_{\alpha\beta} f^{\vphantom\dagger}_{\beta, a, i}.
\end{eqnarray}
Here $\sigma^\mu$ are three spin-1/2 Pauli matrices. Each spinon
$f_{\alpha,a}$ has two indices: $\alpha = \uparrow, \ \downarrow$
denotes spin and $a = 1, 2$ is a ``color'' quantum number. Thus we
can consider not only the usual spin SU(2) rotations in the
$\alpha-\beta$ space, but also color SU(2) transformations in the
$a-b$ space.  Matching with the spin Hilbert space requires not
only constraining the total fermion number to {\it half-filling}
(two fermions per site), but also requiring each site to be an
color SU(2) singlet, which guarantees that on each site the spin
space is a symmetric spin-1 representation:
\begin{eqnarray} && \hat{n}_i = \sum_{a = 1, 2}
  \sum_{\alpha = \uparrow, \downarrow} f^\dagger_{\alpha, a, i}
  f_{\alpha, a, i} = 2,\nonumber \\ && \hat{\tau}^\mu = \sum_{\alpha, a,
    b} f^\dagger_{\alpha, a, i} \tau^\mu_{ab} f_{\alpha, b, i} =
  0. \label{constraint}
\end{eqnarray}
Here $\tau^\mu_{ab}$ are three Pauli matrices that operate on the
color indices. Under time-reversal transformation, in order to
satisfy the commutation relation between Pauli matrices,
$\tau^\mu$ should transform just like spin operators: $\tau^\mu
\rightarrow - \tau^\mu$.

Due to these two independent constraints in Eq.~\ref{constraint},
the spinon $f_{\alpha,a}$ will have a gauge symmetry. In order to
identify the full gauge symmetry, we need to rewrite
$f_{\alpha,a,i}$ in terms of Majorana fermions $\eta$ as follows:
\begin{eqnarray}
  f_{\alpha,a,i} = \frac{1}{2}(\eta_{\alpha,a,1,i} +
  i\eta_{\alpha,a,2,i}) \label{majorana}.
\end{eqnarray}
On every site, $\eta_i$ has in total three two-component spaces,
making the maximal possible transformation on $\eta_i$ SO(8).
Within this SO(8), the spin SU(2) transformations are generated by
the three operators \beqn \vec{\mathbf{S}} = (\sigma^x\lambda^y, \
\ \sigma^y, \ \ \sigma^z \lambda^y ), \label{spin}\eeqn where the
Pauli matrices $\lambda^a$ operate on the two-component space
$(\mathrm{Re}[f], \ \mathrm{Im}[f])$. Under time-reversal
transformation, $\eta \rightarrow \sigma^y \tau^y \lambda^z \eta$,
$i.e.$ both spin and colors are flipped under $\mathbf{Z}^T_2$.

The total gauge symmetry on $\eta$ is the maximal subgroup of
SO(8) that commutes with the spin-SU(2) operators.  This is
$\mathrm{Sp(4)\sim SO(5)}$ generated by the ten matrices
$\Gamma_{ab}=\frac{1}{2i} [\Gamma_a, \Gamma_b]$, where
\begin{eqnarray}
&& \Gamma_{1} = \sigma^y \tau^y \lambda^z, \ \ \Gamma_{2} =
\sigma^y \tau^y \lambda^x, \ \ \Gamma_{3} = \tau^y \lambda^y, \cr
\cr && \Gamma_4 = \tau^x, \ \ \Gamma_5 = \tau^z.
\end{eqnarray}
These $\Gamma^a$ with $a = 1 \cdots 5$ define five gamma matrices
that satisfy the Clifford algebra $\{\Gamma_{a}, \Gamma_b\} =
2\delta_{ab}$. $\Gamma_{ab}$ and $\Gamma_a$ are all $8\times 8$
hermitian matrices. $\Gamma_{ab}$ are all antisymmetric and
imaginary, while $\Gamma_a$ are symmetric.

A spin wave function can be constructed through a slave fermion
wave function, by projecting the mean field ground state to
satisfy the gauge constraints: \begin{eqnarray} |
G_{\mathrm{spin}} \rangle = \prod_i \mathrm{P}(\hat{n}_i =
2)\otimes \mathrm{P}(\hat{\tau}^\mu_i = 0) | f_{\alpha,a} \rangle.
\label{projection}
\end{eqnarray}

Now let us consider the following mean field Hamiltonian of slave
fermion on a {\it honeycomb} lattice: \beqn H_{MF} &=&
\sum_{<i,j>} - t f^\dagger_{i} f_{j} + \sum_{\ll i,j \gg} i\lambda
\nu_{ij} f^\dagger_{i,a} \tau^z_{ab} f_{j,b} \cr\cr &+& \sum_{j} i
t^\prime f^\dagger_{j, a} \tau^x_{ab} f_{j + \sqrt{3}\hat{x}, b} +
H.c. \label{mf} \eeqn Written in terms of the Majorana fermion
$\eta$, the mean field Hamiltonian reads: \beqn H_{MF} &=&
\sum_{<i,j>} - t \eta^t_{i} \Gamma_{12} \eta_{j} + \sum_{\ll i,j
\gg} i\lambda \nu_{ij} \eta^t_{i} \Gamma_5 \eta_{j} \cr\cr &+&
\sum_{j} i t^\prime \eta^t_{j} \Gamma_4 \eta_{j + \sqrt{3}\hat{x}}
+ H.c. \label{mf2} \eeqn

When $t^\prime = 0$, this mean field Hamiltonian is a quantum
``color" Hall Hamiltonian, $i.e.$ it is equivalent to the
Kane-Mele quantum spin Hall
Hamiltonian~\cite{kane2005a,kane2005b}, although instead of a
spin-orbit coupling, in Eq.~\ref{mf} the $\lambda$ term is a
color-orbit coupling. At the mean field level, the quantum color
Hall mean field Hamiltonian Eq.~\ref{mf} has edge states: there is
a left-moving spin-1/2 doublet slave fermion with $\tau^z = 1$,
and another right-moving spin-1/2 doublet slave fermion with
$\tau^z = -1$. Backscattering between left and right moving slave
fermions is forbidden, as long as the time-reversal symmetry and
spin rotation symmetry are preserved.

The $t^\prime$ term is a color-orbit coupling between 2nd neighbor
sites along the $\hat{x}$ directions only. If $t^\prime = 0$, the
color-orbit coupling term $\lambda$ breaks the Sp(4) gauge
symmetry down to its subgroup SO(4), which is generated by
$\Gamma_{ab}$, with $a, b = 1, \cdots 4$; when $t^\prime$ is
nonzero, the gauge symmetry is broken down to SU(2) generated by
$\Gamma_{ab}$, with $a, b = 1, 2, 3$. Since the $t^\prime$ term is
time-reversal even, when $t^\prime$ is small compared with $t$ and
$\lambda$, the edge state cannot be gapped out without degeneracy.
Now the edge states can be describe by the following 1+1d field
theory: \beqn \mathcal{L} = \bar{\eta} \gamma_\mu (\partial_\mu -
i \sum_{l = 1}^3 A^l_\mu G^l ) \eta + \cdots, \label{1dqcd}\eeqn
where $A_\mu^l$ is the residual SU(2) gauge field, and $G^a =
\epsilon_{abc} \Gamma_{bc}$, $a, b , c = 1, 2 ,3$. $\eta$ is the
Majorana fermion introduced in Eq.~\ref{majorana}. $\gamma^0 =
\tau^y$, $\gamma^1 = \tau^x$, $\gamma^5 = \tau^z$, $\bar{\eta} =
\eta^t \gamma^0$.

Eq.~\ref{1dqcd} is precisely the same field theory that describes
the spin-1/2 chain, if we take the standard SU(2) gauge field
formalism for spin-1/2 systems~\cite{wen2002}. There is no
symmetry allowed fermion bilinear terms that can be added to
Eq.~\ref{1dqcd}. 
It is well-known that a spin-1/2 chain must be either gapless or
degenerate, and when it is gapless it can be described by the 1+1d
SU(2)$_1$ conformal field theory (plus marginal perturbations).
Thus we conclude that the boundary of the bulk state Eq.~\ref{mf}
is either a gapless SU(2)$_1$ CFT, or degenerate due to
spontaneous time-reversal symmetry breaking. The spontaneous
time-reversal symmetry breaking can be induced by a relevant four
fermion interaction term in Eq.~\ref{1dqcd}.

It is well-known that the SU(2)$_1$ CFT (and spin-1/2 chain) is
equivalent to an O(4) NLSM with a WZW term (Eq.~\ref{o4wzw}). The
WZW term of Eq.~\ref{o4wzw} can be derived from Eq.~\ref{1dqcd} by
coupling the fermions $\eta$ in Eq.~\ref{1dqcd} to a four
component order parameter $\vec{\phi}$ with unit-length: \beqn
\phi_0 \bar{\eta} \eta + \sum_{k = 1}^3 \phi^k i \bar{\eta}
\gamma^5 \mathbf{S}^k \eta, \label{o4coupling} \eeqn where
$\mathbf{S}^k$ are the three matrices defined in Eq.~\ref{spin}
which generate the spin rotations. Careful analysis shows that the
order parameter $\vec{\phi}$ defined here has the same
transformation as the O(4) vector in Eq.~\ref{o4wzw}.
With the coupling in Eq.~\ref{o4coupling},
after integrating out the slave fermions, a WZW term at precisely
level-1 (Eq.~\ref{o4wzw}) will be generated~\cite{abanov2000}.
Thus the field theory discussed in the previous section can be
precisely derived from this lattice construction. $\phi_0$ changes
sign under time-reversal transformation. If a domain wall of
$\phi_0$ is created, a spin-$1/2$ excitation is localized at the
domain wall, consistent with the field theory discussed below
Eq.~\ref{o3theta} (see also Fig.~\ref{Fig}).

In the bulk the slave fermion is gapped. Since the time-reversal
symmetry in the bulk excludes the existence of a Chern-Simons term
for the residual SU(2) gauge field, this nonabelian gauge field
will lead to confinement, and this confined state has no
topological degeneracy. At the boundary, the effect of this
confinement is not totally understood. For instance, this
confinement might gap out the boundary state through a spontaneous
time-reversal symmetry breaking, namely the order $\langle
\bar{\eta} \eta \rangle \neq 0$ is spontaneously generated. But
nevertheless, the boundary will not be gapped out without
degeneracy.

This lattice construction of SPT can be checked numerically in the
future. Let us define the system on a torus without boundary, and
a spin wave function can be constructed by gauge projecting the
mean field state Eq.~\ref{mf}. Given this projected spin wave
function, one can numerically compute various quantities such as
spin-spin correlation, topological entanglement entropy, and
entanglement spectrum. Since the bulk is completely gapped, the
spin correlation should be short ranged. And since the bulk of the
system has no topological degeneracy, the topological entanglement
entropy, which is defined as a universal constant term in addition
to the standard area law, should be zero. However, since the
system has nontrivial edge states, the edge states should also
exist in the entanglement spectrum, which can be computed using
the projected wave function. We will leave these to future
studies.

\section{4. Discussion and Summary}

In this work we studied a class of two dimensional symmetry
protected topological phases with PSU($N$)$\times \mathbf{Z}^T_2$
symmetry. These SPT phases are described by a 2+1 dimensional
principal chiral model with $\Theta = 2\pi$ (Eq.~\ref{pcm}), and
their boundary states are described by a 1+1d NLSM with
$\Theta^\prime = \pi$, which must be either gapless or degenerate
when the symmetry PSU($N$)$\times \mathbf{Z}^T_2$ is preserved.

The principal chiral model Eq.~\ref{pcm} can describe some other
symmetry protected topological phases as well. For example, the
spin-2 AKLT phase on the square lattice is known to have a
nontrivial 1d edge states. Unlike the 1d Haldane phase, AKLT
states at higher dimensions require translation symmetry to
protect its edge states. For example, on the square lattice, the
edge state of the spin-2 AKLT phase is a spin-1/2 chain, and if
the translation symmetry is broken, this edge spin-1/2 chain will
be dimerized and gapped. The spin-2 AKLT phase on the square
lattice can be viewed as a SPT phase with SO(3)$\times
\mathbf{Z}_2$ symmetry, while here the $\mathbf{Z}_2$ is
translation by one lattice constant instead of time-reversal
transformation. Then Eq.~\ref{o4theta} and Eq.~\ref{pcm} with
$\Theta = 2\pi$ can also describe the two dimensional AKLT phase,
and its generalizations to
arbitrary $N$. 

Our result and analysis apply for all even spatial dimensions. For
large enough $N$ and $m$, $\pi_{2d+1}[\mathrm{SU}(N)] =
\mathbb{Z}$, and $\pi_{2d+1}[\frac{\mathrm{U}(N)}{\mathrm{U}(m)
\times \mathrm{U}(N - m)}] = \mathbb{Z}_1$. Thus at least based on
the field theory, a SPT with PSU($N$)$\times \mathbf{Z}^T_2$
symmetry exists in arbitrary even spatial dimension, and it is
always described by a principal chiral model defined with order
parameter $U$ introduced in Eq.~\ref{order}.

Besides a general classification given in Ref.~\cite{wenspt}, SPT
phases have attracted a lot of attentions and efforts recently
\cite{levinsenthil,luashvin,senthilashvin,lulee,groverashvin}.
Most of these studies are focused on SPT phases with a U(1)
symmetry that is associated with boson number conservation, and
with a U(1) symmetry it is convenient to describe the system using
a dual U(1) gauge field in 2+1d, and a topological response
function of electromagnetic field in 3+1d. The SPT phases
discussed in the current paper and in our previous study
(Ref.~\cite{xu3dspt}) may also have a description in terms of
their responses to external background gauge fields. We will leave
this to future study.

Cenke Xu is supported by the Alfred P. Sloan Foundation, the David
and Lucile Packard Foundation, Hellman Family Foundation, and NSF
Grant No. DMR-1151208. Jeremy Oon is funded by the NSS Scholarship
from the Agency of Science, Technology and Research (A*STAR)
Singapore. Cenke Xu thanks Xiao-Liang Qi for many discussions on
nonlinear sigma models and SPT phases, and thanks Eun-Gook Moon
for insightful questions. Gil Young Cho thanks Joel E. Moore for
encouragement during this work and support from NSF DMR-1206515.


\bibliography{nlsm}

\begin{thebibliography}{31}
\expandafter\ifx\csname natexlab\endcsname\relax\def\natexlab#1{#1}\fi
\expandafter\ifx\csname bibnamefont\endcsname\relax
  \def\bibnamefont#1{#1}\fi
\expandafter\ifx\csname bibfnamefont\endcsname\relax
  \def\bibfnamefont#1{#1}\fi
\expandafter\ifx\csname citenamefont\endcsname\relax
  \def\citenamefont#1{#1}\fi
\expandafter\ifx\csname url\endcsname\relax
  \def\url#1{\texttt{#1}}\fi
\expandafter\ifx\csname urlprefix\endcsname\relax\def\urlprefix{URL }\fi
\providecommand{\bibinfo}[2]{#2}
\providecommand{\eprint}[2][]{\url{#2}}

\bibitem[{\citenamefont{Chen et~al.}(2011)\citenamefont{Chen, Gu, Liu, and
  Wen}}]{wenspt}
\bibinfo{author}{\bibfnamefont{X.}~\bibnamefont{Chen}},
  \bibinfo{author}{\bibfnamefont{Z.-C.} \bibnamefont{Gu}},
  \bibinfo{author}{\bibfnamefont{Z.-X.} \bibnamefont{Liu}}, \bibnamefont{and}
  \bibinfo{author}{\bibfnamefont{X.-G.} \bibnamefont{Wen}},
  \bibinfo{journal}{arXiv:1106.4772}  (\bibinfo{year}{2011}).

\bibitem[{\citenamefont{Liu and Wen}(2012)}]{wenliu}
\bibinfo{author}{\bibfnamefont{Z.-X.} \bibnamefont{Liu}} \bibnamefont{and}
  \bibinfo{author}{\bibfnamefont{X.-G.} \bibnamefont{Wen}},
  \bibinfo{journal}{arXiv:1205.7024}  (\bibinfo{year}{2012}).

\bibitem[{\citenamefont{Haldane}(1983{\natexlab{a}})}]{haldane1}
\bibinfo{author}{\bibfnamefont{F.~D.~M.} \bibnamefont{Haldane}},
  \bibinfo{journal}{Phys. Lett. A} \textbf{\bibinfo{volume}{93}},
  \bibinfo{pages}{464} (\bibinfo{year}{1983}{\natexlab{a}}).

\bibitem[{\citenamefont{Haldane}(1983{\natexlab{b}})}]{haldane2}
\bibinfo{author}{\bibfnamefont{F.~D.~M.} \bibnamefont{Haldane}},
  \bibinfo{journal}{Phys. Rev. Lett.} \textbf{\bibinfo{volume}{50}},
  \bibinfo{pages}{1153} (\bibinfo{year}{1983}{\natexlab{b}}).

\bibitem[{\citenamefont{Ng}(1994)}]{ng1994}
\bibinfo{author}{\bibfnamefont{T.-K.} \bibnamefont{Ng}},
  \bibinfo{journal}{Phys. Rev. B} \textbf{\bibinfo{volume}{50}},
  \bibinfo{pages}{555} (\bibinfo{year}{1994}).

\bibitem[{\citenamefont{Xu}(2012)}]{xu3dspt}
\bibinfo{author}{\bibfnamefont{C.}~\bibnamefont{Xu}},
  \bibinfo{journal}{arXiv:1209.4399}  (\bibinfo{year}{2012}).

\bibitem[{\citenamefont{Read and Sachdev}(1989)}]{sachdev1989nucl}
\bibinfo{author}{\bibfnamefont{N.}~\bibnamefont{Read}} \bibnamefont{and}
  \bibinfo{author}{\bibfnamefont{S.}~\bibnamefont{Sachdev}},
  \bibinfo{journal}{Nucl. Phys. B} \textbf{\bibinfo{volume}{316}},
  \bibinfo{pages}{609} (\bibinfo{year}{1989}).

\bibitem[{\citenamefont{Read and Sachdev}(1990)}]{sachdev1990}
\bibinfo{author}{\bibfnamefont{N.}~\bibnamefont{Read}} \bibnamefont{and}
  \bibinfo{author}{\bibfnamefont{S.}~\bibnamefont{Sachdev}},
  \bibinfo{journal}{Phys. Rev. B} \textbf{\bibinfo{volume}{42}},
  \bibinfo{pages}{4568} (\bibinfo{year}{1990}).

\bibitem[{\citenamefont{Witten}(1984)}]{witten1984}
\bibinfo{author}{\bibfnamefont{E.}~\bibnamefont{Witten}},
  \bibinfo{journal}{Commun. Math. Phys.} \textbf{\bibinfo{volume}{92}},
  \bibinfo{pages}{455} (\bibinfo{year}{1984}).

\bibitem[{\citenamefont{Knizhnik and
  Zamolodchikov}(1984)}]{KnizhnikZamolodchikov1984}
\bibinfo{author}{\bibfnamefont{V.~G.} \bibnamefont{Knizhnik}} \bibnamefont{and}
  \bibinfo{author}{\bibfnamefont{A.~B.} \bibnamefont{Zamolodchikov}},
  \bibinfo{journal}{Nucl. Phys. B} \textbf{\bibinfo{volume}{247}},
  \bibinfo{pages}{83} (\bibinfo{year}{1984}).

\bibitem[{\citenamefont{lan Affleck}(1985)}]{affleck1985}
\bibinfo{author}{\bibnamefont{lan Affleck}}, \bibinfo{journal}{Nucl. Phys. B}
  \textbf{\bibinfo{volume}{257}}, \bibinfo{pages}{397} (\bibinfo{year}{1985}).

\bibitem[{\citenamefont{Lieb et~al.}(1961)\citenamefont{Lieb, Schultz, and
  Mattis}}]{LSM}
\bibinfo{author}{\bibfnamefont{E.~H.} \bibnamefont{Lieb}},
  \bibinfo{author}{\bibfnamefont{T.~D.} \bibnamefont{Schultz}},
  \bibnamefont{and} \bibinfo{author}{\bibfnamefont{D.~C.}
  \bibnamefont{Mattis}}, \bibinfo{journal}{Ann. Phys.}
  \textbf{\bibinfo{volume}{16}}, \bibinfo{pages}{407} (\bibinfo{year}{1961}).

\bibitem[{\citenamefont{Coleman}(1976)}]{coleman1976}
\bibinfo{author}{\bibfnamefont{S.}~\bibnamefont{Coleman}},
  \bibinfo{journal}{Ann. of Phys.} \textbf{\bibinfo{volume}{101}},
  \bibinfo{pages}{239} (\bibinfo{year}{1976}).

\bibitem[{\citenamefont{Levine et~al.}(1983)\citenamefont{Levine, Libby, and
  Pruisken}}]{pruisken1}
\bibinfo{author}{\bibfnamefont{H.}~\bibnamefont{Levine}},
  \bibinfo{author}{\bibfnamefont{S.~B.} \bibnamefont{Libby}}, \bibnamefont{and}
  \bibinfo{author}{\bibfnamefont{A.~M.~M.} \bibnamefont{Pruisken}},
  \bibinfo{journal}{Phys. Rev. Lett.} \textbf{\bibinfo{volume}{51}},
  \bibinfo{pages}{1915} (\bibinfo{year}{1983}).

\bibitem[{\citenamefont{Levine et~al.}(1984{\natexlab{a}})\citenamefont{Levine,
  Libby, and Pruisken}}]{pruisken2}
\bibinfo{author}{\bibfnamefont{H.}~\bibnamefont{Levine}},
  \bibinfo{author}{\bibfnamefont{S.~B.} \bibnamefont{Libby}}, \bibnamefont{and}
  \bibinfo{author}{\bibfnamefont{A.~M.~M.} \bibnamefont{Pruisken}},
  \bibinfo{journal}{Nucl. Phys. B} \textbf{\bibinfo{volume}{240}},
  \bibinfo{pages}{30, 49, 71} (\bibinfo{year}{1984}{\natexlab{a}}).

\bibitem[{\citenamefont{Levine et~al.}(1984{\natexlab{b}})\citenamefont{Levine,
  Libby, and Pruisken}}]{pruisken3}
\bibinfo{author}{\bibfnamefont{H.}~\bibnamefont{Levine}},
  \bibinfo{author}{\bibfnamefont{S.~B.} \bibnamefont{Libby}}, \bibnamefont{and}
  \bibinfo{author}{\bibfnamefont{A.~M.~M.} \bibnamefont{Pruisken}},
  \bibinfo{journal}{Nucl. Phys. B} \textbf{\bibinfo{volume}{240}},
  \bibinfo{pages}{49} (\bibinfo{year}{1984}{\natexlab{b}}).

\bibitem[{\citenamefont{Levine et~al.}(1984{\natexlab{c}})\citenamefont{Levine,
  Libby, and Pruisken}}]{pruisken4}
\bibinfo{author}{\bibfnamefont{H.}~\bibnamefont{Levine}},
  \bibinfo{author}{\bibfnamefont{S.~B.} \bibnamefont{Libby}}, \bibnamefont{and}
  \bibinfo{author}{\bibfnamefont{A.~M.~M.} \bibnamefont{Pruisken}},
  \bibinfo{journal}{Nucl. Phys. B} \textbf{\bibinfo{volume}{240}},
  \bibinfo{pages}{71} (\bibinfo{year}{1984}{\natexlab{c}}).

\bibitem[{\citenamefont{Pruisken et~al.}(2001)\citenamefont{Pruisken, Baranov,
  and Voropaev}}]{pruisken2011}
\bibinfo{author}{\bibfnamefont{A.~M.~M.} \bibnamefont{Pruisken}},
  \bibinfo{author}{\bibfnamefont{M.~A.} \bibnamefont{Baranov}},
  \bibnamefont{and} \bibinfo{author}{\bibfnamefont{M.}~\bibnamefont{Voropaev}},
  \bibinfo{journal}{arXiv:cond-mat/0101003}  (\bibinfo{year}{2001}).

\bibitem[{\citenamefont{Xu and Ludwig}(2011)}]{xuludwig}
\bibinfo{author}{\bibfnamefont{C.}~\bibnamefont{Xu}} \bibnamefont{and}
  \bibinfo{author}{\bibfnamefont{A.~W.~W.} \bibnamefont{Ludwig}},
  \bibinfo{journal}{arXiv:1112.5303}  (\bibinfo{year}{2011}).

\bibitem[{\citenamefont{Wu et~al.}(2006)\citenamefont{Wu, Bernevig, and
  Zhang}}]{wuedge}
\bibinfo{author}{\bibfnamefont{C.}~\bibnamefont{Wu}},
  \bibinfo{author}{\bibfnamefont{B.~A.} \bibnamefont{Bernevig}},
  \bibnamefont{and} \bibinfo{author}{\bibfnamefont{S.-C.} \bibnamefont{Zhang}},
  \bibinfo{journal}{Phys. Rev. Lett.} \textbf{\bibinfo{volume}{96}},
  \bibinfo{pages}{106401} (\bibinfo{year}{2006}).

\bibitem[{\citenamefont{Senthil and Fisher}(2005)}]{senthilfisher}
\bibinfo{author}{\bibfnamefont{T.}~\bibnamefont{Senthil}} \bibnamefont{and}
  \bibinfo{author}{\bibfnamefont{M.~P.~A.} \bibnamefont{Fisher}},
  \bibinfo{journal}{Phys. Rev. B} \textbf{\bibinfo{volume}{74}},
  \bibinfo{pages}{064405} (\bibinfo{year}{2005}).

\bibitem[{\citenamefont{Xu et~al.}(2012)\citenamefont{Xu, Wang, Qi, Balents,
  and Fisher}}]{xuspin1}
\bibinfo{author}{\bibfnamefont{C.}~\bibnamefont{Xu}},
  \bibinfo{author}{\bibfnamefont{F.}~\bibnamefont{Wang}},
  \bibinfo{author}{\bibfnamefont{Y.}~\bibnamefont{Qi}},
  \bibinfo{author}{\bibfnamefont{L.}~\bibnamefont{Balents}}, \bibnamefont{and}
  \bibinfo{author}{\bibfnamefont{M.~P.~A.} \bibnamefont{Fisher}},
  \bibinfo{journal}{Phys. Rev. Lett.} \textbf{\bibinfo{volume}{108}},
  \bibinfo{pages}{087204} (\bibinfo{year}{2012}).

\bibitem[{\citenamefont{Kane and Mele}(2005{\natexlab{a}})}]{kane2005a}
\bibinfo{author}{\bibfnamefont{C.~L.} \bibnamefont{Kane}} \bibnamefont{and}
  \bibinfo{author}{\bibfnamefont{E.~J.} \bibnamefont{Mele}},
  \bibinfo{journal}{Physical Review Letter} \textbf{\bibinfo{volume}{95}},
  \bibinfo{pages}{226801} (\bibinfo{year}{2005}{\natexlab{a}}).

\bibitem[{\citenamefont{Kane and Mele}(2005{\natexlab{b}})}]{kane2005b}
\bibinfo{author}{\bibfnamefont{C.~L.} \bibnamefont{Kane}} \bibnamefont{and}
  \bibinfo{author}{\bibfnamefont{E.~J.} \bibnamefont{Mele}},
  \bibinfo{journal}{Physical Review Letter} \textbf{\bibinfo{volume}{95}},
  \bibinfo{pages}{146802} (\bibinfo{year}{2005}{\natexlab{b}}).

\bibitem[{\citenamefont{Wen}(2002)}]{wen2002}
\bibinfo{author}{\bibfnamefont{X.-G.} \bibnamefont{Wen}},
  \bibinfo{journal}{Phys. Rev. B} \textbf{\bibinfo{volume}{65}},
  \bibinfo{pages}{165113} (\bibinfo{year}{2002}).

\bibitem[{\citenamefont{Abanov and Wiegmann}(2000)}]{abanov2000}
\bibinfo{author}{\bibfnamefont{A.~G.} \bibnamefont{Abanov}} \bibnamefont{and}
  \bibinfo{author}{\bibfnamefont{P.~B.} \bibnamefont{Wiegmann}},
  \bibinfo{journal}{Nucl. Phys. B} \textbf{\bibinfo{volume}{570}},
  \bibinfo{pages}{685} (\bibinfo{year}{2000}).

\bibitem[{\citenamefont{Levin and Senthil}(2012)}]{levinsenthil}
\bibinfo{author}{\bibfnamefont{M.}~\bibnamefont{Levin}} \bibnamefont{and}
  \bibinfo{author}{\bibfnamefont{T.}~\bibnamefont{Senthil}},
  \bibinfo{journal}{arXiv:1206.1604}  (\bibinfo{year}{2012}).

\bibitem[{\citenamefont{Lu and Vishwanath}(2012)}]{luashvin}
\bibinfo{author}{\bibfnamefont{Y.-M.} \bibnamefont{Lu}} \bibnamefont{and}
  \bibinfo{author}{\bibfnamefont{A.}~\bibnamefont{Vishwanath}},
  \bibinfo{journal}{Phys. Rev. B} \textbf{\bibinfo{volume}{86}},
  \bibinfo{pages}{125119} (\bibinfo{year}{2012}).

\bibitem[{\citenamefont{Ashvin~Vishwanath}(2012)}]{senthilashvin}
\bibinfo{author}{\bibfnamefont{T.~S.} \bibnamefont{Ashvin~Vishwanath}},
  \bibinfo{journal}{arXiv:1209.3058}  (\bibinfo{year}{2012}).

\bibitem[{\citenamefont{Lu and Lee}(2012)}]{lulee}
\bibinfo{author}{\bibfnamefont{Y.-M.} \bibnamefont{Lu}} \bibnamefont{and}
  \bibinfo{author}{\bibfnamefont{D.-H.} \bibnamefont{Lee}},
  \bibinfo{journal}{arXiv:1210.0909}  (\bibinfo{year}{2012}).

\bibitem[{\citenamefont{Grover and Vishwanath}(2012)}]{groverashvin}
\bibinfo{author}{\bibfnamefont{T.}~\bibnamefont{Grover}} \bibnamefont{and}
  \bibinfo{author}{\bibfnamefont{A.}~\bibnamefont{Vishwanath}},
  \bibinfo{journal}{arXiv:1210.0907}  (\bibinfo{year}{2012}).

\end{thebibliography}

\end{document}